\documentclass[dvipsnames,aps,prl,nobibnotes,twocolumn,superscriptaddress,nolongbibliography]{revtex4-1}

\usepackage[T1]{fontenc}
\usepackage[english]{babel}
\usepackage[utf8]{inputenc}

\usepackage[unicode=true,pdfusetitle,bookmarks=true,bookmarksnumbered=false,bookmarksopen=false,
 breaklinks=false,pdfborder={0 0 0},pdfborderstyle={},backref=false,colorlinks=true]
 {hyperref}
 \hypersetup{
 citecolor=blue,
 urlcolor=blue,
 linkcolor=blue}

\usepackage{array}
\usepackage{wrapfig}
\usepackage{float}
\usepackage{units}
\usepackage{amsmath, relsize}
\usepackage{cancel}
\usepackage{mathrsfs}
\usepackage{amssymb}
\usepackage{graphicx}
\usepackage{stackrel}
\usepackage{babel}
\usepackage{xparse}
\usepackage{soul}
\usepackage[final]{pdfpages}
\usepackage{comment}
\usepackage{tcolorbox}
\usepackage{nicefrac}

\makeatletter
\AtBeginDocument{\let\LS@rot\@undefined}
\makeatother

\begin{document}
\title{The Synergistic Route to Stretched Criticality}

\author{Lorenzo Lucarini}
\affiliation{Physics Department and INFN, University of Rome Tor Vergata, 00133 Rome, Italy}
\affiliation{``Enrico Fermi'' Research Center (CREF), Via Panisperna 89A, 00184 - Rome, Italy}
\author{Sandro Meloni}
\affiliation{Institute for Cross-Disciplinary Physics and Complex Systems (IFISC), CSIC-UIB, 07122 - Palma de Mallorca, Spain}
\affiliation{``Enrico Fermi'' Research Center (CREF), Via Panisperna 89A, 00184 - Rome, Italy}
\author{Pablo Villegas}
\email[]{pablo.villegas@cref.it}
\affiliation{``Enrico Fermi'' Research Center (CREF), Via Panisperna 89A, 00184 - Rome, Italy}
\affiliation{Instituto Carlos I de F\'isica Te\'orica y Computacional, Univ. de Granada, E-18071, Granada, Spain.}

\begin{abstract}
Griffiths phases are typically associated with quenched disorder, while frustration gives rise to multistability and spin-glass behavior. Whether extended criticality can arise in other contexts remains an open question. Here, we show that synergistic interactions provide a distinct route to non-conventional critical phenomena. By combining spreading mechanisms that reinforce activity through complementary pathways, we uncover a broad distribution of relaxation rates, leading to Griffiths-like slow dynamics and extended criticality. We demonstrate that this mechanism is robust across networks and emerges both in systems with explicit higher-order interactions and in purely pairwise systems with nonlinear dynamical rules.
\end{abstract}

\maketitle

Exotic collective behavior in complex systems is fundamentally shaped by the structure of their interactions~\cite{HenkelBook}. In statistical physics, different forms of microscopic constraints are known to generate non-conventional macroscopic dynamical regimes. For instance, quenched disorder in interaction strengths leads to \emph{Griffiths phases}~\cite{Griffiths1969,Vojta2006}, where rare regions with locally favorable conditions sustain activity for exponentially long times. The combination of exponentially rare clusters with exponentially long survival times produces anomalously slow relaxation and generic power-law decay of activity~\cite{Munoz2010}. Extended critical regions have been proposed as a mechanism to relax the fine-tuning problem of criticality in complex systems, replacing isolated transition points with broad regimes of critical-like dynamics in complex networks~\cite{Moretti2013,RMP_Munoz,Odor2015}. 
Similarly, antagonistic or frustrated interactions give rise to \emph{spin-glass} behavior~\cite{Mezard1987,Binder1986}. In this case, rugged energy landscapes and multiple metastable states lead to intrinsically slow, history-dependent dynamics, including aging and memory effects~\cite{Mezard1987,facchetti2011computing}. A crucial question that remains open is whether further mechanisms can lead to non-ergodic behavior.

Many real-world phenomena, however, are dominated neither by disorder nor by antagonism, but by cooperative interactions~\cite{levin2014public}. In biological, social, and technological systems, groups of agents often act synergistically, producing outcomes that cannot be reduced to independent pairwise interactions~\cite{petri2014homological,Centola2010,Sanjuan2017,LuciaSanz2017}. For instance, symbiogenesis and competition jointly shape Darwinian dynamics~\cite{sagan1967origin}. In viral infections, for example, infectivity can depend nonlinearly on the local multiplicity of infection due to cooperative replication of multiple viral genomes~\cite{AndreuMoreno2020}. Similarly, interactions between pathogens shape epidemic outbreaks, giving rise to unexpected phenomena, including explosive transitions or hysteresis~\cite{sanz2014,Cai2015,Cui2019,costa2022,Ghanbarnejad2022,lamata2024}. Such cooperative mechanisms can arise either through explicit higher-order interactions~\cite{iacopini2019}  
or through nonlinear activation rules requiring the joint influence of multiple neighbors~\cite{Granovetter1978,SorianoPanos2019,Bizhani2012,Grassberger2016}. This diversity of phenomena raises the question of whether synergy provides a potential route to stretched criticality.


In this Letter, we show that the coupling between distinct spreading mechanisms generates extended critical regimes. Their cooperative reinforcement produces a broad distribution of relaxation rates, leading to Griffiths-like slow decay. This behavior can be explained by an accumulation of low-lying spectral modes. We also demonstrate that this mechanism is robust across network structures and arises both in systems with explicit higher-order interactions and in purely pairwise dynamics. Cooperative reinforcement thus provides a fundamental route to broad relaxation spectra and extended critical behavior.


\begin{figure*}[hbtp]
    \centering
    \includegraphics[width=0.8\textwidth]{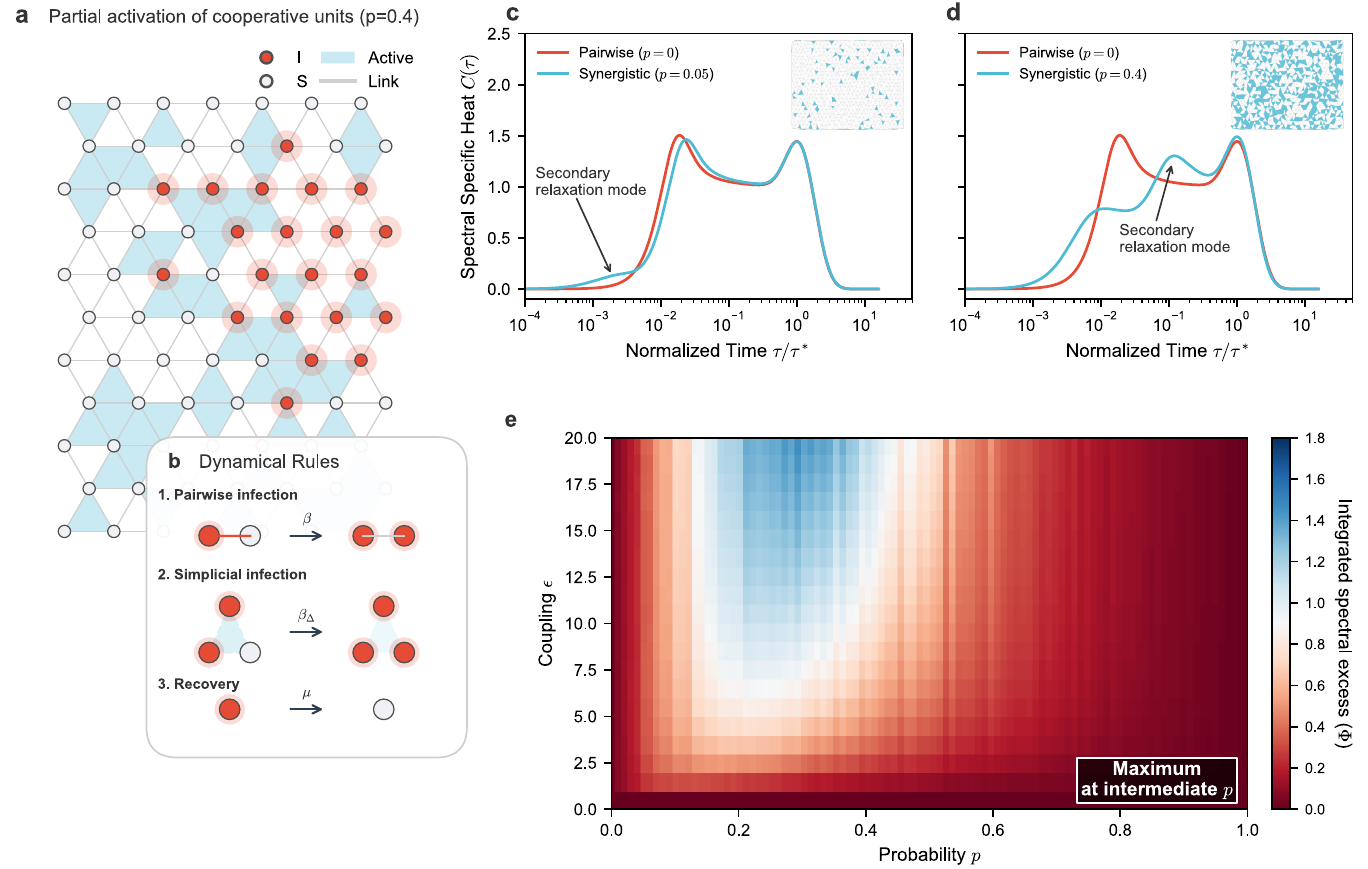}
    \caption{\label{fig:simplicial_dynamics}
    \textbf{Partial activation of cooperative units generates additional slow relaxation modes.}
    (a) Triangular lattice with an active fraction $p$ of 2-simplices (triangles, teal) superimposed on the pairwise backbone. Node colors represent SIS states: susceptible (white) and infected (red). 
    (b) Spreading dynamics combining pairwise infection ($\beta$), cooperative simplicial infection ($\beta_{\Delta}$), and recovery ($\mu$). 
    (c),(d) Spectral specific heat $C$ as a function of normalized diffusion time $\tau/\tau^*$ for two values of the active simplicial fraction, $p=0.05$ and $p=0.4$, for $N=1023$. Relative to the pairwise baseline ($p=0$, red), partial activation of simplices generates an additional peak at short times, signaling the emergence of a secondary relaxation mode. Insets show the corresponding spatial arrangement of active triangles. 
    (e) Phase diagram of the integrated spectral excess $\Phi$ in the $(p,\epsilon)$ plane for $\epsilon=10$, and $N=1023$. The main difference is concentrated in an intermediate region of $p$, indicating that the effect is strongest when cooperative motifs are neither absent nor fully saturated. Note also the absence of spectral excess for low coupling, $\epsilon$.}
\end{figure*}

\textit{Laplacian analysis of synergistic mechanisms} --- To isolate the structural origin of synergistic slowing down, we consider the simplest setting in which a pairwise diffusive substrate is coupled to a partially activated set of simplicial units. Specifically, we first analyze a triangular lattice in which a fraction $p$ of 2-simplices (i.e., triangles) is activated, interpolating between a purely pairwise ($p=0$) and a fully saturated simplicial structure ($p=1$), as illustrated in Figure~\ref{fig:simplicial_dynamics}(a).

To describe the corresponding relaxation backbone, we introduce a composite Laplacian operator
\begin{equation}
\hat{L}_{\mathrm{syn}}=\hat{L}^{(1)}+\epsilon\,\hat{P}\circ \hat{L}^{(2)},
\label{eq:composite_L}
\end{equation}
where $\hat{L}^{(1)}=\hat{D}-\hat{A}$ is the standard graph Laplacian, $\hat{L}^{(2)}$ is the generalized Laplacian associated with 2-simplices, $\epsilon$ controls the relative weight of the higher-order contribution, and $\hat{P}$ selects the active fraction $p$ of triangles. The operator $\hat{L}^{(2)}$ is constructed from the generalized simplicial adjacency $\hat{A}^{(2)}$ and degree $\hat{K}^{(2)}$ through $\hat{L}^{(2)}_{ij}=2\hat{K}^{(2)}_i\delta_{ij}-\hat{A}^{(2)}_{ij}$~\cite{Max2022}. In this way, $p$ controls the geometry of cooperative reinforcement, while $\epsilon$ controls the synergistic strength.

We analyze the resulting multiscale organization through the diffusion density operator \cite{Manlio2016,Villegas2022}, $\hat{\rho}(\tau)=\frac{e^{-\tau \hat{L}_{\mathrm{syn}}}}{\mathrm{Tr}\,e^{-\tau \hat{L}_{\mathrm{syn}}}}$,
where $\tau$ is the diffusion time. From $\hat{\rho}(\tau)$ we compute the von Neumann entropy $S(\tau)=-\mathrm{Tr}\,[\hat{\rho}(\tau)\log \hat{\rho}(\tau)]$, and its logarithmic derivative, $C(\tau)=-\frac{dS}{d\log \tau}$, which, acting as a spectral specific heat, reveals the characteristic relaxation scales of the lattice~\cite{Villegas2022,Villegas2023}, as shown in Figure~\ref{fig:simplicial_dynamics}(c).

Finally, to quantify the strength of additional spectral modes, we define the integrated spectral excess, namely,
\begin{equation}
\Phi =\int_{0}^{\infty}
\left[
C^{(syn)}(\tau/\tau_\infty^*)-C^{(0)}(\tau/\tau_\infty^*)
\right]\, d\log\tau,
\label{eq:Order}
\end{equation}
where $\tau_\infty^*$ is the slowest relaxation timescale of the system, $C^{(syn)}(\tau/\tau_\infty^*)$ is the temporally rescaled specific heat for the composite Laplacian described in Eq.~\eqref{eq:composite_L}, and $C^{(0)}(\tau/\tau_\infty^*)$ corresponds to the curve with $p=0$ (i.e., to the 2D triangular lattice). By construction, $\Phi$ measures the total spectral weight that is generated by cooperative couplings beyond the pairwise reference. The resulting phase diagram in the $(p,\epsilon)$ plane [see Figure~\ref{fig:simplicial_dynamics}(e)] reveals a finite intermediate region in which $\Phi$ is large, showing that the effect is strongest when cooperative reinforcement is neither absent nor fully saturated.

For the purely pairwise case ($p=0$ or $\epsilon=0$), $C(\tau)$ displays the expected plateau of a regular two-dimensional substrate, $C\simeq d_s/2$~\cite{Poggialini2025} [see Figure~\ref{fig:simplicial_dynamics}(c)]. Instead, for intermediate values of $p$, increasing interaction complexity gives rise to a nontrivial effect: the emergence of an intermediate regime, driven by the partial activation of cooperative units, where the spectral analysis reveals additional relaxation scales. Remarkably, the fully saturated limit $p=1$ collapses onto the same functional form, independently of $\epsilon$, i.e., recovering the diffusion properties of the pairwise case. As reported in Figures~\ref{fig:simplicial_dynamics}(c,d), partial simplicial activation generates an additional peak at short times, absent in the pairwise baseline, signaling the emergence of a secondary relaxation mode.

The resulting excess of low-lying modes provides a structural signature of the synergistic regime and suggests a mechanism for the slow dynamics observed below.

\textit{Stretched criticality via higher-order synergistic interactions.}--- To probe the dynamical consequences of the additional relaxation mode, we explore a simplicial Susceptible-Infected-Susceptible (SIS) process~\cite{iacopini2019}. In the model, contagion occurs through two processes on top of the 2D triangular lattice [see Figures~\ref{fig:simplicial_dynamics}(a,b)]: pairwise infections at rate $\beta$ along links, and higher-order infections at rate $\beta_{\Delta}$ within the fraction of active 2-simplices, selected by \(P\), while recovery takes place at rate $\mu$. The ratio $\beta_\Delta/\beta$ is expected to play a role analogous to the coupling $\epsilon$ introduced above.

\begin{figure}[hbtp]
    \centering
    \includegraphics[width=1.0\linewidth]{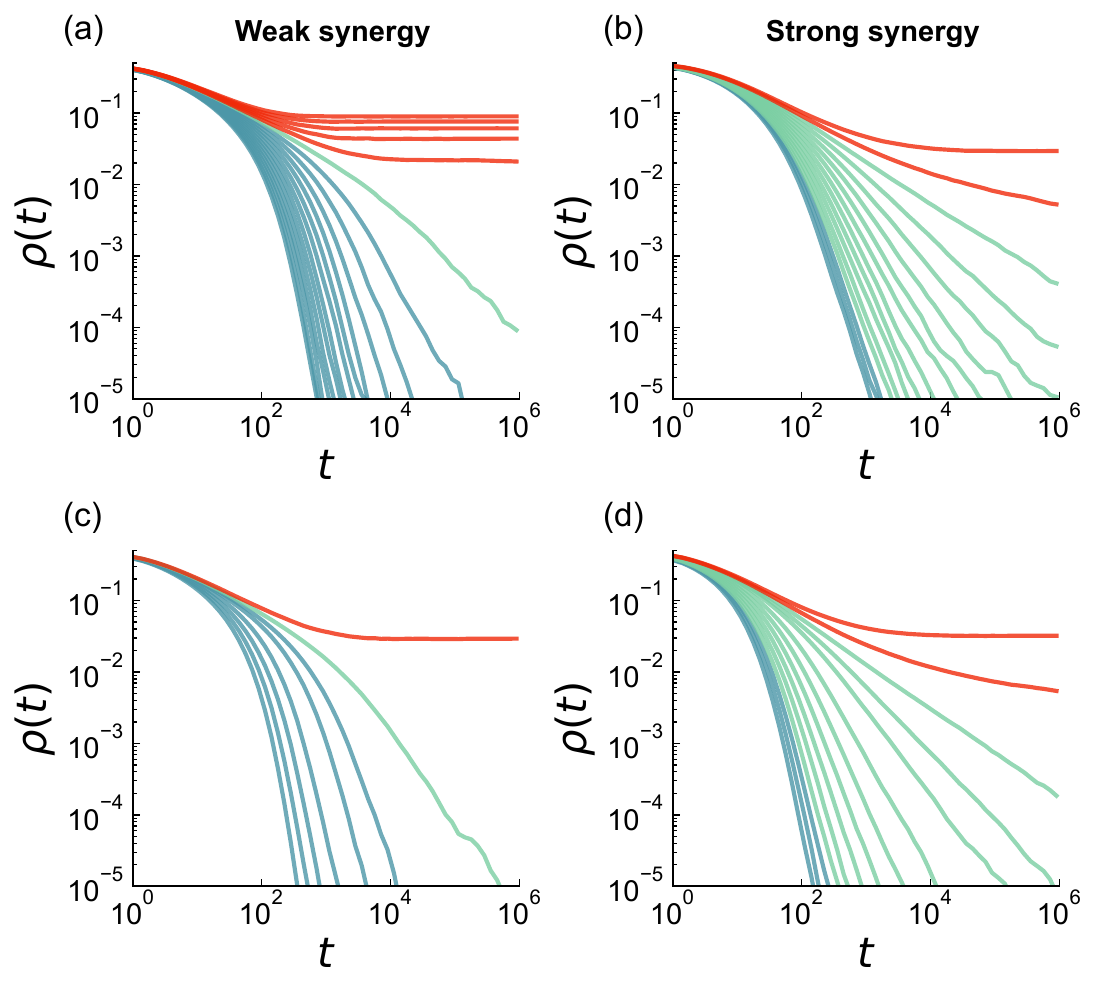}
    \caption{
\textbf{Extended region of slow relaxation induced by synergistic interactions.}
Temporal evolution of the infection density $\rho(t)$ for two values of the cooperative parameter, $p=0.05$ (weak synergy) and $p=0.25$ (strong synergy).
(a,b) Simplicial SIS dynamics: (a) $\beta \in (0.220,0.238)$ with $\beta_c=0.233$; (b) $\beta \in (0.150,0.165)$ with $\beta_c^{low}\equiv\beta_{c}^{l}=0.152$ and $\beta_c^{high}\equiv\beta_{c}^{h}=0.163$.
(c,d) Pairwise quadratic contact process: (c) $\beta \in (0.220,0.242)$ with $\beta_c=0.240$; (d) $\beta \in (0.140,0.184)$ with $\beta_c^l=0.152$ and $\beta_c^h=0.180$.
In both models, green curves highlight the broad region of algebraic-like decay over several decades, signaling the emergence of an extended critical regime associated with a broad spectrum of relaxation times. Blue curves correspond to the absorbing phase and red curves to the active phase.
Parameters: $\beta_\Delta/\beta=10$, $\mu=1$, $N=2^{17}$. Curves are averaged over $10^3$ realizations.}  
    \label{fig:Ho-GP}
\end{figure}

Figures~\ref{fig:Ho-GP}(a,b) report the temporal decay of the infection density $\rho(t)$ for parameter values lying in the regime where the spectral excess $\Phi$ is low ($p=0.05$) and maximal ($p=0.25$), while maintaining a large ratio $\beta_\Delta/\beta$ [see Figure~\ref{fig:simplicial_dynamics}(e) and SM~\cite{SM}]. In the latter case, instead of a conventional transition separating exponential decay from sustained activity, the system develops a broad region in which $\rho(t)$ decays algebraically over several decades. The effective decay exponent varies continuously across this region, indicating the absence of a single critical point and the emergence of an extended critical regime. By contrast, for low values of the ratio $\beta_\Delta/\beta$, no extended slow regime is observed, being consistent with the behavior of the spectral excess (see SM~\cite{SM}). Therefore, these results are consistent with the previously hypothesized connection between stretched criticality and the aforementioned redistribution of Laplacian eigenmodes toward low frequencies.

This behavior originates from the coexistence of spreading processes with distinct relaxation scales: one provides the propagation backbone, while the other locally reinforces activity through nonlinear transmission. Their synergistic interplay generates a broad relaxation spectrum through the superposition of many exponential modes~\cite{Vojta2006}. 
This gives rise to rare-region-like dynamical domains with anomalously long lifetimes~\cite{Munoz2010,Vojta2006}.

\textit{Pairwise realization of synergistic criticality.}--- We now demonstrate that explicit higher-order structure is not required for the emergence of synergistically reinforced extended criticality. Previous work has established that higher-order contagion processes can be mapped onto pairwise nonlinear dynamics, where higher-order terms are equivalent to facilitation effects~\cite{Meloni2025}. Motivated by this correspondence, we consider a pairwise SIS model with dual dynamics \cite{Bottcher, Ohtsuki1987, Villa2014,Elgart2006}. The baseline is the standard contact process, in which infection spreads at a rate $\beta$. In addition, the fraction $p$ now corresponds to nodes that also follow a threshold activation rule: a susceptible node becomes infected at rate $\beta_{\Delta}$ only when exposed to at least two infected neighbors (e.g, the $S + 2I \to 3I$ reaction). 

\begin{figure*}[hbtp]
    \centering
    \includegraphics[width=0.8\linewidth]{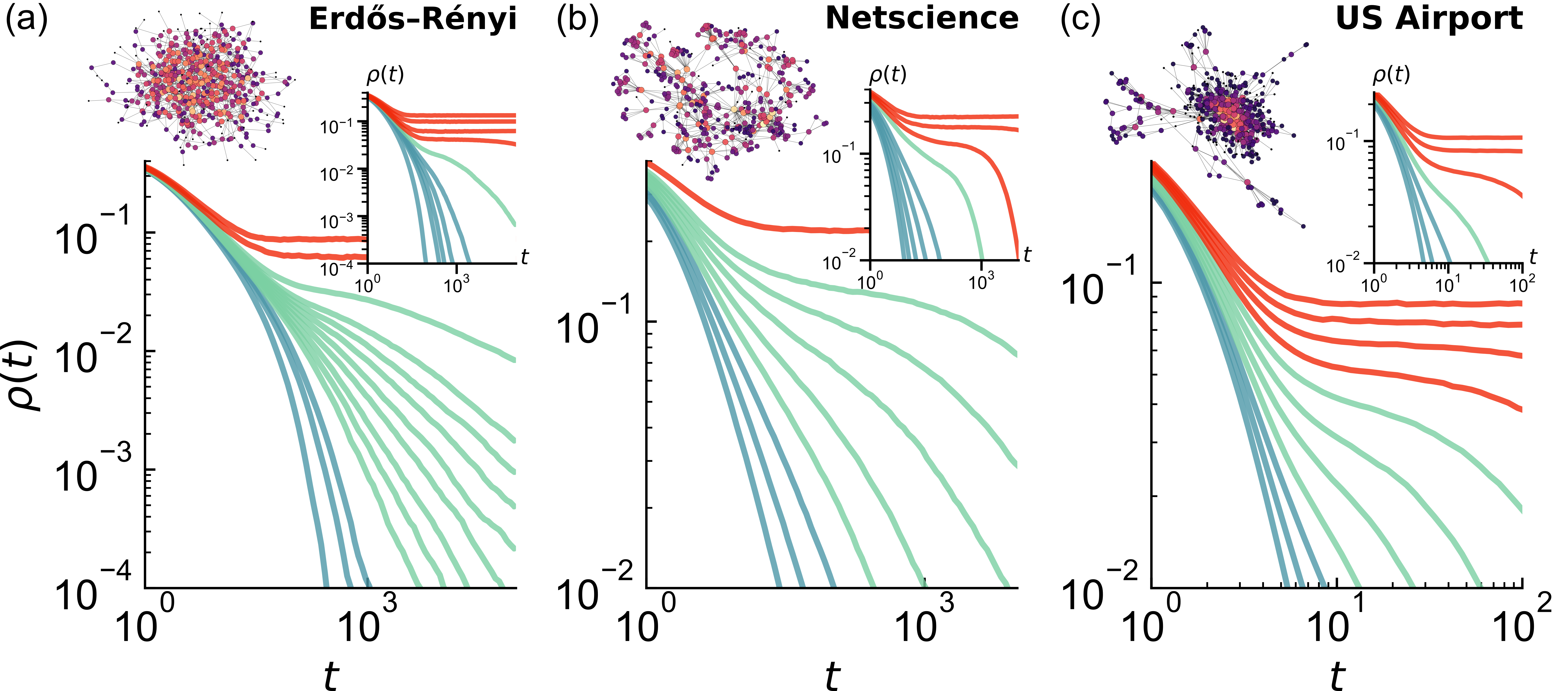}
    \caption{
\textbf{Robustness across network structures.}
Temporal evolution of the infection density $\rho(t)$ for the pairwise cooperative model on different network topologies for strong and zero synergy (insets, $p=0$). Blue curves correspond to the absorbing phase and red curves to the active one, while the green curves highlight algebraic-like decay. (a) Erd\H{o}s--R\'enyi network ($N=10^{4}$, $\langle \kappa \rangle = 3$, $p=0.10$), with $\beta \in (0.31,0.35)$, $\beta_c^l=0.32$, and $\beta_c^h=0.33$. Inset: $\beta \in (0.30,0.40)$, with $\beta_c=0.340$.
(b) Scientific collaboration network (Netsci, $N=379$, $p=0.30$), with $\beta \in (0.10,0.20)$, $\beta_c^l=0.13$, and $\beta_c^h=0.17$. Inset: $\beta \in (0.10,0.28)$, with $\beta_c=0.22$.
(c) US airport network ($N=500$, $p=0.20$), with $\beta \in (0.01,0.03)$, $\beta_c^l=0.016$, and $\beta_c^h=0.022$. Inset: $\beta \in (0.01,0.04)$, with $\beta_c=0.025$. 
In all cases, the system develops a broad region of slow, algebraic-like decay, indicating that the emergence of stretched criticality is robust across heterogeneous and real-world network structures, and is not tied to lattice geometry. Parameters: $\beta_\Delta/\beta=10$, $\mu=1$. Curves are averaged over $10^3-10^4$ realizations.}
    \label{fig:CNets-Syn}
\end{figure*}

Figures~\ref{fig:Ho-GP}(c,d) show that the purely pairwise system exhibits the same phenomenology as the simplicial case: a broad region of slow, algebraic-like decay and continuously varying relaxation rates for intermediate $p$-values.

To test the robustness of this mechanism beyond regular lattices, we study the same pairwise dynamics on synthetic and real networks. Figure~\ref{fig:CNets-Syn} shows the temporal evolution of the infection density on Erdős–Rényi networks, a scientific collaboration network~\cite{Newman2006}, and the US airport network~\cite{Colizza2007}. In all cases, the system develops a broad region of slow, algebraic-like decay. While finite-size effects and network heterogeneity may prevent a perfectly clean power-law behavior, as illustrated in the US airport network, synergistic spreading consistently enhances slow relaxation near criticality, broadening the dynamical regime over which activity exhibits anomalously slow decay.

These results show that the mechanism is not tied to a specific representation (lattice geometry, or higher-order interactions), but instead persists across synthetic and real-world networks as a general consequence of synergistic dynamics. This bridges to realistic epidemic settings, where cooperative effects arise on finite, heterogeneous contact networks rather than idealized substrates, without requiring fine-tuning to a single epidemic threshold.

\textit{Outlook.}--- Our results identify the synergistic interplay between distinct spreading mechanisms as a minimal and general route to stretched criticality. In particular, activity is reinforced through complementary dynamical pathways acting on different timescales, producing a broad spectrum of relaxation times and Griffiths-like slow dynamics over extended parameters regions. Unlike Griffiths phases induced by structural heterogeneity of the underlying network~\cite{Moretti2013,Odor2015} (e.g., degree heterogeneity), where slow dynamics arise from static node-to-node variability in local spreading conditions, here the broad relaxation spectrum emerges dynamically from synergistic infections. Because the same phenomenology appears both in simplicial spreading and in purely pairwise quadratic activation, the effect is not tied to a specific higher-order structure but reflects a general consequence of synergistic dynamics. In this sense, synergy acts as a dynamical organizing principle that reshapes the relaxation spectrum of the system.

More broadly, our findings show that rare-region-like behavior can arise from dynamical constraints associated with cooperation~\cite{Vojta2006,Munoz2010}. Across the examples studied here, structural heterogeneity alone does not generate evident slow dynamics, which instead emerges from the synergistic reinforcement between standard pairwise transmission and nonlinear infections. More generally, synergy provides a distinct microscopic route to stretched criticality (non-ergodicity), complementing the well-established roles of structural disorder and frustration in statistical physics. Such mechanisms may extend beyond spreading dynamics to systems where competing interactions or constraints generate multiple intrinsic scales, such as frustrated functional materials with topologically protected structures~\cite{Falsi2025}. The resulting regimes expand the parameter space where slow collective dynamics arise, suggesting that a broad class of systems may not require fine-tuning to a single critical point.

From a biological perspective, our results may be particularly relevant in systems where activation requires collective or repeated exposure, such as viral infections with multiplicity-dependent infectivity~\cite{AndreuMoreno2020} or cooperative epidemics~\cite{lansbury2020,henrich2025}. In these contexts, synergistic spreading may effectively implement forms of dynamical bet-hedging~\cite{Kussell2005}, allowing systems to balance rapid spreading with long-lived persistence through a broad distribution of timescales. More generally, the interplay between local infection and network-mediated spreading may provide a systematic framework to understand how multi-agent and multi-species systems sustain activity across scales.

\paragraph{\textbf{Acknowledgments.}}
S.M. acknowledges the Spanish State Research Agency (MICIU/AEI/10.13039/501100011033) and FEDER (UE) under projects COSASTI (PID2024-157493NB-C22), and the Mar{\'\i}a de Maeztu project CEX2021-001164-M.
P.V. acknowledges the Spanish Ministry and Agencia Estatal de Investigaci\'on (AEI), MICIN/AEI/10.13039/501100011033, for financial support, Project PID2023-149174NB-I00 funded also by
ERDF/EU. P.V. thanks T.~Gili, M.~Lucas, G.~Petri, and F.~Battiston for early discussions that helped shape some of the ideas developed in this work.
\newpage

\def\url#1{}
%

\clearpage


\section*{Supplementary material}

# Supplementary Information: The Synergistic Route to Stretched Criticality

## CONTENTS

## I. ACTIVITY DECAY FOR DIFFERENT COOPERATIVE DENSITIES AND COUPLINGS

To characterize how synergistic interactions reshape the dynamical regime, we systematically analyze the temporal decay of the activity density $\rho(t)$ across different values of the cooperative fraction $p$ and infection-rate ratio $\beta_\Delta/\beta$. While the main text focuses on representative cases near the region of maximal spectral excess, here we provide a broader view of the phenomenology across parameter space through an extensive analysis of the simplicial cooperative model.

Our goal is twofold: (i) to quantify how the extent of the slow-relaxation regime depends on the density of cooperative 2-simplices, and (ii) to clarify how the interplay between pairwise and simplicial infection channels controls the emergence of stretched criticality.

### A. High sinergy region

We first consider regimes with a large ratio $\beta_\Delta/\beta$, where cooperative interactions are dynamically dominant (high-$p$), a regime not shown in the main manuscript.

As shown in Fig. S1, the system exhibits an extended region of slow, algebraic-like decay up to intermediate cooperative densities, $p \simeq 0.5$, in agreement with the spectral analysis reported in the main text. In this regime, cooperative effects generate a broad distribution of relaxation times, leading to stretched critical behavior.

As $p$ increases, however (e.g., $p = 0.75$, Fig. S1(b)), the phenomenology changes qualitatively. The system exhibits a much sharper transition between decay and sustained activity, suggestive of a first-order-like transition. This reflects the fact that, when cooperative motifs become sufficiently dense, the system effectively homogenizes at the level of dynamical reinforcement. As a result, the coexistence of distinct relaxation pathways is suppressed, and the broad spectrum of relaxation times collapses toward the classical behavior.

This crossover indicates that stretched criticality is maximized at intermediate cooperative densities and is progressively lost as the system approaches the fully cooperative limit.

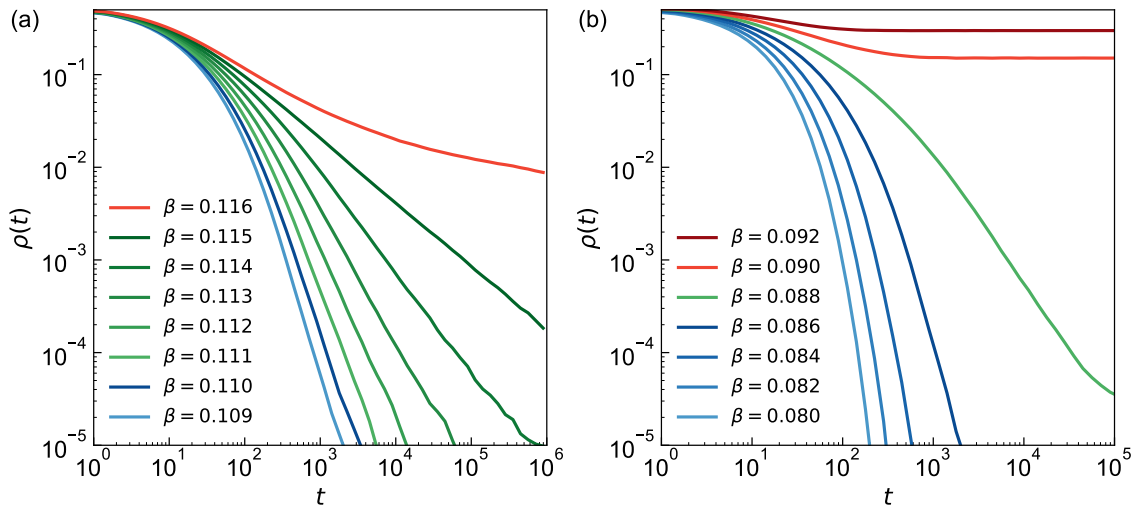

FIG. S1. **Activity decay in the high cooperative-density regime.** Temporal evolution of the infection density $\rho(t)$ for large fractions of cooperative 2-simplices: (a) $p = 0.50$, (b) $p = 0.75$. Each curve corresponds to a different value of the infection rate $\beta$, at fixed $\beta_\Delta/\beta = 10$. At intermediate cooperative density ($p = 0.50$), the system exhibits extended slow decay (green curves), indicating a broad distribution of relaxation times. In contrast, at higher density ($p = 0.75$), the transition becomes sharper, and the slow-relaxation regime is strongly reduced, signaling a crossover toward a homogeneous dynamical behavior. Red curves correspond to the active phase, while the blue lines correspond to the absorbing one. Parameters: $N = 2^{17}$. All curves have been averaged over $10^2 - 5 \cdot 10^3$ independent realizations.

## B. Low cooperative-coupling regime

We now turn to the regime of weak cooperative reinforcement, corresponding to low values of the ratio $\beta_\Delta/\beta$. In this limit, the simplicial infection channel is not strong enough to generate the broad distribution of relaxation times underlying stretched criticality. This is consistent with the spectral excess analysis, $\Phi(p,\epsilon)$, presented in the main text [see Fig. 1(e)].

Figure S2 shows the temporal decay of the activity density $\rho(t)$ for different values of $p$ and $\beta$ in this weak-cooperation regime. In contrast to the cases discussed in the main text and in the previous subsection, the system does not display an extended region of algebraic-like decay. Instead, for all values of $p$, the decay curves are consistent with a conventional continuous transition between an absorbing phase and an active phase.

This behavior indicates that varying the fraction of available 2-simplices alone is not sufficient to produce stretched criticality: a sufficiently strong cooperative channel is also required. When $\beta_\Delta/\beta$ is small, cooperative events act only as a perturbative correction to the pairwise dynamics. As a result, the coexistence of distinct relaxation pathways is too weak to generate the broad relaxation spectrum observed in the stretched-critical regime.

These results further support the interpretation that stretched criticality emerges only within an intermediate region of cooperative density and cooperative strength, where pairwise contagion and simplicial reinforcement operate on comparable timescales.

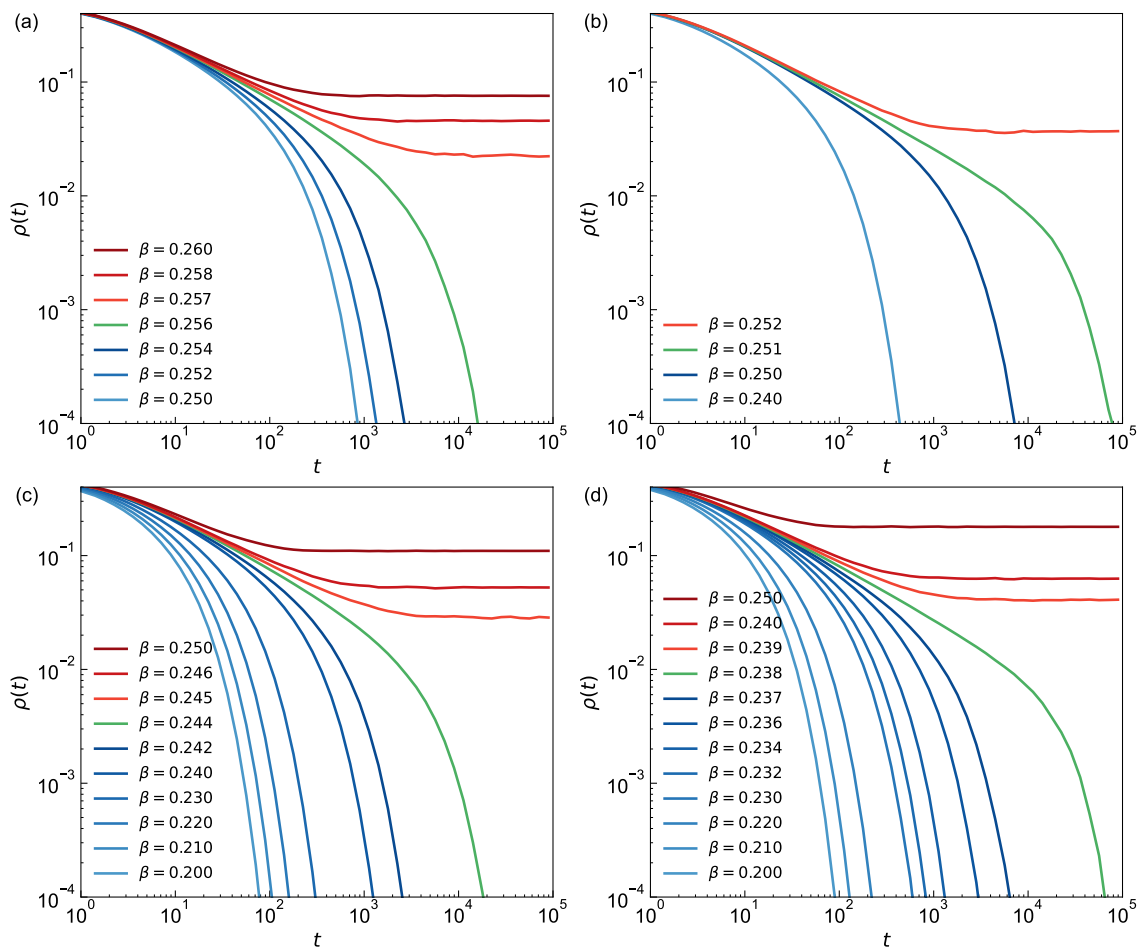

FIG. S2. **Activity decay in the weak-cooperation regime.** Temporal evolution of the infection density $\rho(t)$ for low values of the ratio $\beta_\Delta/\beta = 0.5$ (a) $p = 0.05$, (b) $p = 0.25$, (c) $p = 0.50$, and (d) $p = 0.75$. Each curve corresponds to a different value of the pairwise infection rate $\beta$. For all values of $p$, the system exhibits a conventional continuous transition between decay and sustained activity, with no evidence of an extended algebraic-like regime. Red curves correspond to the active phase, the green line to the critical point, while the blue lines correspond to the absorbing phase. This shows that weak synergistic effects are insufficient to produce stretched criticality. Parameters: $N = 2^{17}$. All curves have been averaged over $10^2 - 5 \cdot 10^3$ independent realizations.



\end{document}